\documentclass[10pt,showpacs,twocolumn]{revtex4}
\usepackage{graphicx}
\usepackage{amsmath}
\usepackage{amssymb}
\usepackage{}
\begin{document}
\title{Revisiting the tunnelling site of electrons in strong field enhanced ionization of molecules}
\author{Cheng Huang, Pengfei Lan\footnote
{Corresponding author: pengfeilan@hust.edu.cn}, Yueming Zhou,
Qingbin Zhang, Kunlong Liu, and Peixiang Lu\footnote {Corresponding
author: lupeixiang@mail.hust.edu.cn}}

\affiliation{School of Physics and Key Laboratory of Fundamental
Physical Quantities Measurement of Ministry of Education, Huazhong
University of Science and Technology, Wuhan 430074, People's
Republic of China}
\date{\today}

\begin{abstract}
We investigated electron emissions in strong field enhanced
ionization of asymmetric diatomic molecules by quantum calculations.
It is demonstrated that the widely-used intuitive physical picture,
i.e., electron wave packet direct ionization from the up-field site
(DIU), is incomplete. Besides DIU, we find another two new
ionization channels, the field-induced excitation with subsequent
ionization from the down-field site (ESID), and the up-field site
(ESIU). The contributions from these channels depend on the
molecular asymmetry and internuclear distance. Our work provides a
more comprehensive physical picture for the long-standing issue
about enhanced ionization of diatomic molecules.

\end{abstract}
 \maketitle

Tunnelling ionization is one of the most fundamental quantum effects
when atoms and molecules are exposed to strong laser field. As the
doorway step of various strong-field processes, such as, high-order
harmonic and attosecond pulse generation \cite{Hentschel,Krausz},
double ionization \cite{Fittingoff,Becker} and high-order
above-threshold ionization \cite{Paulus,Becker2}, understanding the
ionization dynamics is of essential importance for controlling the
electron dynamics in these processes. Moreover, molecular ionization
signal itself also preserves some information of the molecular
structure, and thus can be used to image molecular structure
\cite{DPavi,Kamta2}. Therefore, the ionization has attracted
significant interests over the past several decades. Theories,
including PPT \cite{Perelomov}, ADK \cite{Ammosov}, have been well
established for atoms. Lots of efforts have also been made to extend
these theories to molecules \cite{Tong}. Nevertheless, because the
molecules have more degrees of freedom and more complicated
structure, the underlying physics becomes richer and the ionization
dynamics is still not completely clear yet. It has been demonstrated
that when the molecule is stretched to a critical internuclear
distance $R_c$ the ionization probability sharply increases, which
is called enhanced ionization (EI)
\cite{Codling,Seideman,Zuo,Kamta,Schmidt,Normand,Constant,Gibson}.
An intuitive physical picture \cite{Codling,Seideman,Zuo} based on
the quasi-static tunneling theory \cite{Delone} have been proposed
to explain the behavior of molecular EI. When the molecular is
stretched to the critical distance $R_c$, an inner potential barrier
between the two cores emerges and localizes the electron population
at each of cores. Then, the up-field population only needs to tunnel
through the inner barrier directly to the continuum, which is
considerably easier than tunnelling through the outer barrier
between the down-field core and the continuum. Thus a remarkable
enhancement of the ionization probability happens around the
critical distance $R_c$. According to the intuitive physical
picture, electron wave packet direct ionization from the up-field
site (DIU) is considered responsible for molecular EI.

Although such a DIU physical picture has been commonly used to
analyze and explain the experiments of molecular ionization and
related processes \cite{Pavi}, the physical picture of molecular EI
is still unclear and confusing. For instance, in Ref. \cite{Betsch},
Betsch{\it et al.} measured the ejection direction of multiply
charged ion fragments from a variety of molecules (N$_2$, O$_2$, CO,
CO$_2$ and HBr) driven by a two-color laser field. The observed
forward-backward dissociation asymmetries imply that the electron is
preferentially emitted from the down-field site, in contradiction
with the DIU physical picture. Recently, a single-color elliptically
polarized laser pulse is used to probe the tunnelling site of
electrons from the dimer ArXe by angular streaking technique
\cite{Wu,Eckle}. Wu {\it et al.} reported that the ionization more
easily happens at the up-field site, supporting the DIU physical
picture. Because the intuitive physical picture is based on the
quasi-static theory, lacking a perspective on the dynamics of
ionization processes, controversy still exists in these experiments.

To understand the long-standing issues about EI
\cite{Betsch,Wu,Sheehy,Thompson,Ray}, in this Letter, we investigate
the electron dynamics and the tunnelling site by carefully examining
time evolution of the electron density and ionization rate with
numerically solving time-dependent Schr\"{o}dinger equation (TDSE).
A more comprehensive physical picture is established for EI dynamics
of diatomic molecules. Besides the DIU ionization channel, we find
another two new ionization channels. The contributions from these
channels depend on the asymmetry and internuclear distance of the
molecules.

\begin{figure}
\begin{center}
\includegraphics[width=9.0cm,clip]{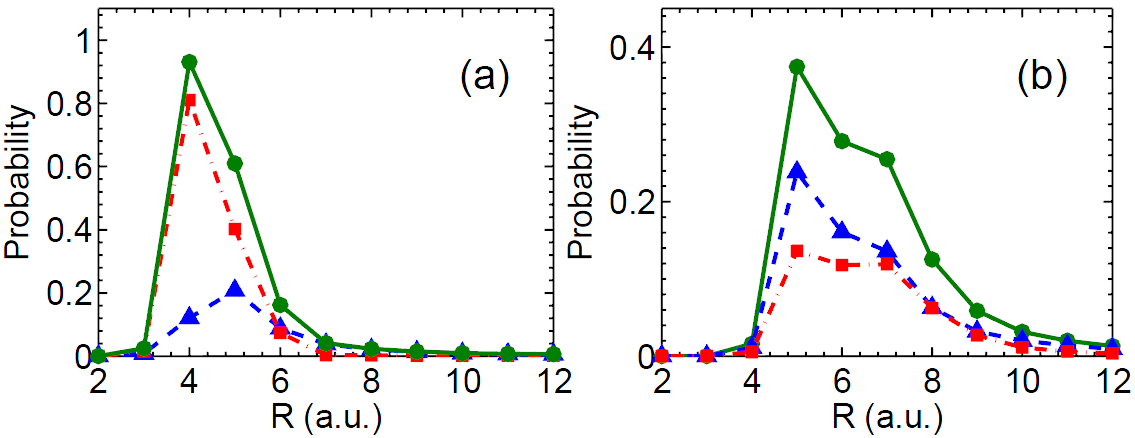}
\caption{\label{fig1} (color online) Total (green), Left (red), and
Right (blue) ionization probabilities as a function of internuclear
distance R. (a) the molecule with large asymmetry; (b) the molecule
with small asymmetry.}
\end{center}
\end{figure}

This work is intended to explore a general effect, rather than to
model a special experiment, so we consider a generic model diatomic
molecule aligned along the electric field vector of the linearly
polarized light. The two-dimensional TDSE can be written as [atomic
units (a.u.) are used throughout this paper unless stated
otherwise]:
$i\frac{\partial\psi(x,y,t)}{\partial{t}}=H(x,y,t)\psi(x,y,t),$
where x, y denote the electron coordinates. H(x,y,t) is the
Hamiltonian and reads
$H(x,y,t)=[-\frac{1}{2}(\frac{\partial^{2}}{\partial{x}^{2}}
+\frac{\partial^{2}}{\partial{y}^{2}})-\frac{Z_1}{\sqrt{(x+R/2)^2+y^2+a}}
-\frac{Z_2}{\sqrt{(x-R/2)^2+y^2+b}}+xE(t)]$. R is the internuclear
distance. $Z_1$, $Z_2$ are the electric charges of two nuclei, which
are fixed at (-R/2,0) and (+R/2,0), respectively. a, b are the
screening parameters of left and right nuclei.
E(t)=E$_0$$\sin$($\pi$t/$\tau_p$)$^2$$\cos$($\omega$t) is the
electric field of the laser pulse, with the angular frequency
$\omega$=0.057 a.u. (corresponding to the wavelength 800 nm) and the
full duration $\tau_p$=10T (T is the laser cycle). In order to
investigate the role of molecular asymmetry, we chose a set of
parameters $Z_1$=2, $Z_2$=1, a=0.5 and b=0.5 to represent a model
molecule with large asymmetry (e.g. HeH$^{2+}$). The asymmetry is
defined by the parameter A=I$_{pl}$/I$_{pr}$, where I$_{pl}$ and
I$_{pr}$ denote the ionization energies of the left and the right
cores when the neighboring core is removed, respectively. According
to this definition, the asymmetry parameter is A=1.38/0.54=2.6. The
other set of parameters $Z_1$=1, $Z_2$=1, a=0.39 and b=0.92 is used
to represent a model molecule with small asymmetry (e.g. ArXe$^+$).
Its asymmetry parameter is A$=0.58/0.45=1.3$. In our work the laser
intensity of 1$\times$10$^{15}$ W/cm$^2$ is used for the former
molecule, and 9$\times$10$^{13}$ W/cm$^2$ for the latter molecule.
The split-operator spectral method \cite{Feit} is used to
numerically solve the TDSE.

\begin{figure}
\begin{center}
\includegraphics[width=7.0cm,clip]{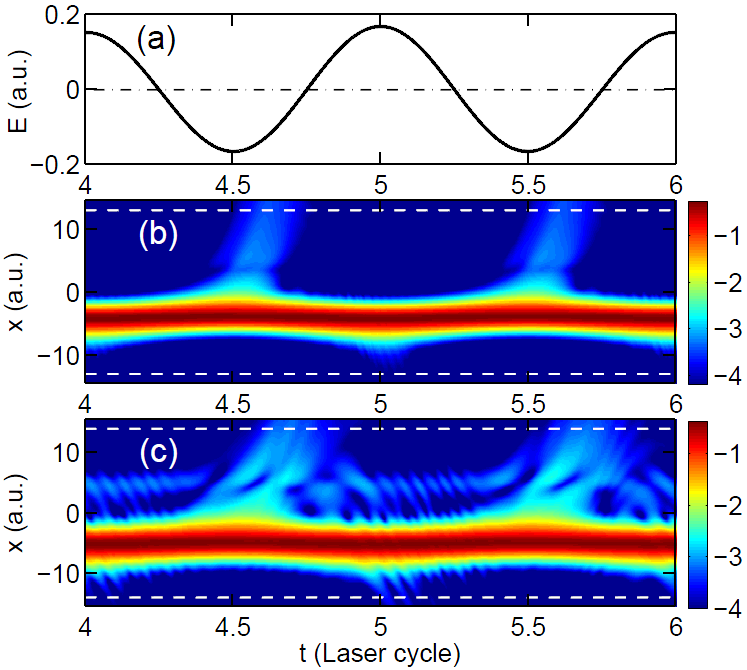}
\caption{\label{fig2} (color online) (a): The laser electric field.
(b) and (c): Electron density as a function of time and the
coordinate x for the molecule with large asymmetry at R=8 a.u. and
for the molecule with small asymmetry at R=10 a.u.}
\end{center}
\end{figure}

Figure 1(a) and (b) show the ionization probabilities as a function
of internuclear distance R for the molecules with large and small
asymmetries, respectively. The red and blue curves show the
ionization probabilities from the left ($x<0$) and right ($x>0$)
sides, which are obtained by integrating the probability flux at
x=-R/2-9 and x=R/2+9 from the beginning to end of the laser pulse.
The green curve represents the total ionization probability. With
the increase of R, the total ionization probabilities for these two
molecules both firstly increase and then gradually decrease. A
remarkable enhancement happens around $R=4$ a.u. and 6 a.u. for the
molecule with large asymmetry [see Fig. 1(a)] and small asymmetry
[see Fig. 1(b)], respectively. However, one can see a distinct
difference between these two molecules. For the molecule with large
asymmetry, the probability of electrons escaping from the left side
is much larger than that from the right side around the critical
distance. Whereas for the molecule with small asymmetry, the
ionization probability from the left side is slightly smaller than
that from the right side around the critical distance. At large
internuclear distance (R$>8$), both molecules show slightly more
electrons emitted from the right side.

In order to explore if the electron is emitted from the up-field
site or down-field site, we carefully examine the time evolution of
the electron density along the polarization direction. We first
discuss the EI at large internuclear distance. Figure 2(b) and 2(c)
show the electron density as a function of time for the molecule
with large asymmetry at R=8 a.u. and for the molecule with small
asymmetry at R=10 a.u. respectively. Recall that the electron is
preferentially emitted from the right side at these internuclear
distances as shown in Fig. 1. From Fig. 2(b) and 2(c) one can see
that the ionization mainly occurs at the two instants around t=4.5T
and t=5.5T. At those times, the electric field is negative and thus
the left core is up-field. The result indicates that electron wave
packet located at the left (i.e., up-field) core directly tunnels
through the inner potential barrier to the continuum. This
ionization channel, so-called direct ionization from the up-field
site (DIU), is consistent with the intuitive physical picture of
molecular EI. Therefore, for asymmetric diatomic molecules DIU is
the dominant ionization channel at the large internuclear distance.

\begin{figure}
\begin{center}
\includegraphics[width=8.0cm,clip]{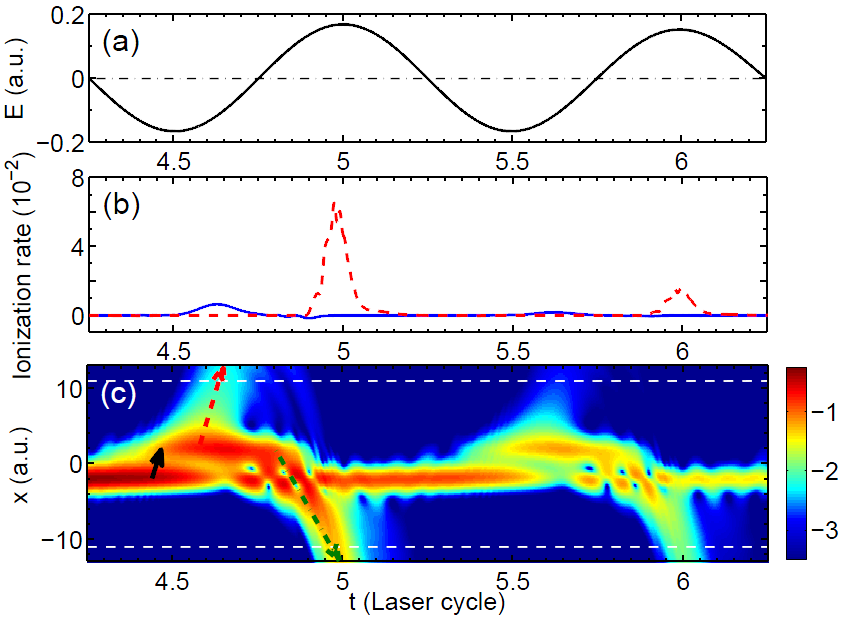}
\caption{\label{fig2} (color online) (a) The laser electric field.
(b) Ionization rate from the left (red) and right (blue) sides as a
function of time. (c) Electron density as a function of time and the
coordinate x. Molecule with large asymmetry at R=4 a.u.}
\end{center}
\end{figure}

Next, we discuss the ionization dynamics at the relatively small
internuclear distance. Figure 3(b) shows the ionization rate from
the left (red curve) and right (blue curve) sides as a function of
time for the molecule with large asymmetry at R=4 a.u. One can see
that the dominant ionization burst is from the left side around
t=5.0T, when the electric field is positive. There are also some
electron wave packets escaping away from the right side with low
probabilities around t=4.65T and t=5.65T, and from the left side
around t=6.0T. In order to more clearly reveal the dynamics of
electron emissions, the time evolution of the electron density is
examined. As shown in Fig. 3(c), the molecule is initially at the
ground state and the electron wave packet is dominantly localized at
the left core. At the instant of t=4.45T, some electron population
is firstly excited to the right core \cite{Kamta}, as indicated by
the black arrow. A short time later, at the instant of t=4.6T a
small part of the excited population leaves from the right core [see
the red arrow]. When this electron wave packet arrives at x=11 a.u.
[the white dashed curve], it is considered that ionization occurs.
At this time the electric field is still negative and thus the right
core is down-filed. That is to say, the electron escapes away from
the down-field site by this process. Furthermore, more excited
population remains localized at the right core. When the electric
field becomes positive and the right core is promoted to the
up-field site, the excited population quickly tunnels through the
inner potential barrier to the continuum around t=5.0T [see the
green arrow], which corresponds to the highest ionization peak in
Fig. 3(b). In this channel the electron is emitted from the up-field
site. Different from DIU channel at the large R, the ionization
channel at small R mentioned above is a two-step process. The first
step is that the electron population located at the left core is
excited to the right core when the electric field is negative. Then
the excited electron wave packet can be ionized by two paths. One
path is that the excited electron wave packet tunnels through the
right outer potential barrier to ionize from the down-field site
when the electric field is negative. The other path is that the
excited electron wave packet stays until the electric field
reverses, and then goes through the inner potential barrier to
directly ionize from the up-field site. Moreover, for large
asymmetric molecules, there is more excited electron population
ionized from the up-field site.

\begin{figure}
\begin{center}
\includegraphics[width=8.0cm,clip]{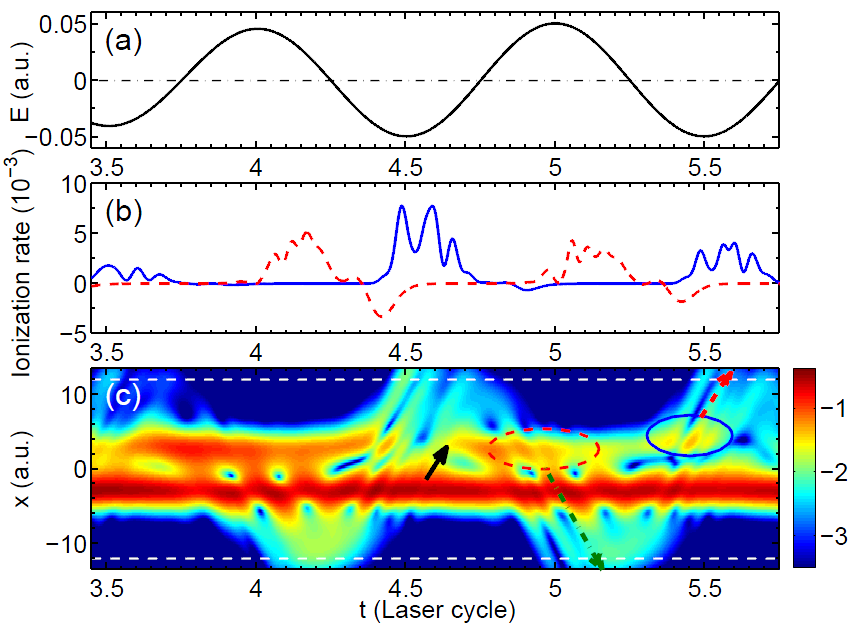}
\caption{\label{fig4} (color online) (a) The laser electric field.
(b) Ionization rate from the left (red) and right (blue) sides as a
function of time. (c) Electron density as a function of time and the
coordinate x. Molecule with small asymmetry at R=6 a.u.}
\end{center}
\end{figure}

Further, we analyze the ionization dynamics for small asymmetric
molecules. Figure 4 shows the ionization rate from the left (red
curve) and right (blue curve) sides and the electron density along
the polarization direction as a function of the time for the
molecule with small asymmetry at R=6 a.u. Due to the periodicity of
ionization signal, we only need to analyze the region of 4.6T-5.7T.
One can see that a part of electron population is excited to the
right core at t=4.6T [see the black arrow]. Then the electric field
turns positive at t=4.75T and within the subsequent positive
half-cycle [4.75T,5.25T] a part of excited electron population
tunnels through the inner barrier to ionize from the left side [see
the green arrow]. After the electric field reverses again at
t=5.25T, the right core is lowered to the down-field site. The
residual excited population localized at the right core tunnels
through the right outer barrier to the continuum around t=5.5T [see
the red arrow]. Similar to the molecule with large asymmetry, these
two ionization channels are also a two-step process. The only
difference is that those excited electron population emitted from
the down-field site stays at the right core for a longer time. The
emitted electrons from the left and right sides correspond to the
ionization of the up-field and down-field sites, respectively.
Furthermore, we integrate the ionization rate from the left and the
right sides shown in Fig. 4(b) over the time. The result reveals
that the ionization probability from the right side is slightly
larger than that from the left side. That is to say, the excited
electron population is more likely ionized from the down-field site,
which is opposite to the case of the molecule with large asymmetry
and also in contradiction with the DIU physical picture. This result
indicates that the tunneling site in EI depends on the molecular
asymmetry.

In addition, as compared with the molecule with large asymmetry, the
ionization rate curves for the molecule with small asymmetry are
wider, as shown in Fig. 4(b). Moreover, one can see
multiple peaks in the ionization rate cures. The
similar multiple-peak structure within a half-cycle of the laser
field have also been found for the H$_2^+$ in Ref.
\cite{Takemoto} and attributed to the transient localization
of the electron at one of the nuclei \cite{Kawata}. In Fig. 4(c) the similar
transient localization of the electron is also visible and results
in the wider time distribution of ionization signal.

\begin{figure}
\begin{center}
\includegraphics[width=7.0cm,clip]{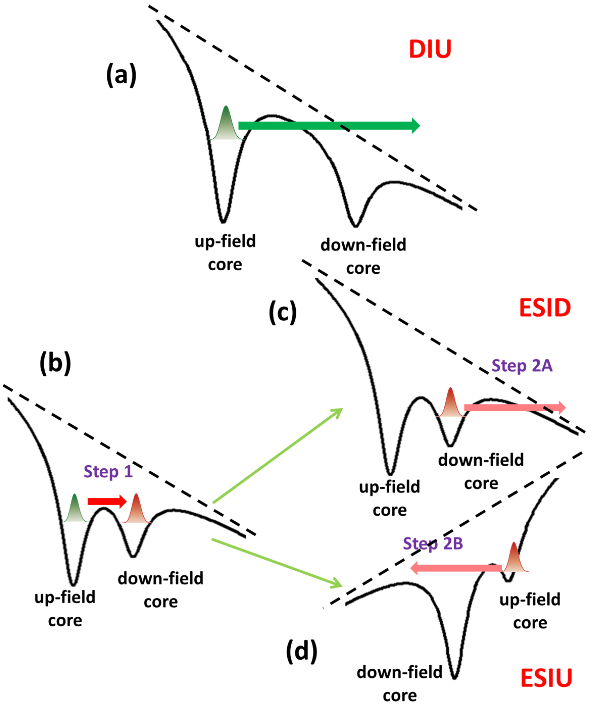}
\caption{\label{fig3} (color online) Sketches of the three different
ionization channels.}
\end{center}
\end{figure}

Our results suggest that the following scenario takes place for EI
of diatomic molecules: There are three main ionization channels, as
shown in Fig. 5. At the large internuclear distance, the electron
located at the left core directly tunnels through the inner
potential barrier between the two cores to the continuum, as shown
in Fig. 5(a). This ionization channel, electron wave packet direct
ionization from the up-field site (DIU), is consistent with the
intuitive physical picture for the molecular EI. As the internuclear
distance decreases the contribution from the DIU channel quickly
decreases. At the small internuclear distance, the other two
ionization channels dominate. Both of the two channels are a
two-step process, and their first step is the same. The first step
is that the electron population located at the left core is excited
to the right core when the electric field is negative [see Fig.
5(b)]. Then the excited electron wave packet can be emitted by two
paths. One path is that the excited electron wave packet around the
right core tunnels through the right outer barrier to the continuum
when the electric field is negative. In this case the right core is
down-field [see Fig. 5(c)]. So this ionization channel can be called
field-induced excitation with subsequent ionization from the
down-field site (ESID). The other path is that the excited electron
wave packet stays until the electric field turns positive. Then the
excited electron wave packet tunnels through the inner potential
barrier directly to the continuum. This ionization channel is
referred to as field-induced excitation with subsequent ionization
from the up-field site (ESIU), as shown in Fig. 5(b) and 5(d).

In conclusion, we have investigated the dynamics of electron
emissions in strong field EI of diatomic molecules by numerically
solving TDSE. It is found that there are three ionization channels
leading to ionization enhancement. Their relative contributions are
related to the molecular asymmetry and internuclear distance. At the
large internuclear distance the dominant contribution is from DIU
ionization channel regardless of molecular asymmetry, which is
consistent with the intuitive physical picture of EI. However, at
small internuclear distance the other two new ionization channels
dominate and their relative contributions depend on the molecular
asymmetry. For the molecule with large asymmetry the electron is
preferentially ionized from the up-field site by the ESIU channel.
Whereas for the molecule with small asymmetry the electron is more
likely ionized from the down-field site by the ESID channel. Our
work provides a more comprehensive physical picture for EI of
diatomic molecules. It can promote the understanding of the
dissociation dynamics of molecules.

This work was supported by the National Natural Science Foundation
of China under Grant No. 61275126, 11234004, and the 973 Program of
China under Grant No. 2011CB808103.

\end{document}